\documentclass[pre,eqsecnum,showpacs,byrevtex,amsmath,amssymb,nofootinbib]{revtex4}
 \usepackage{amsmath,graphicx}

\begin{document}

\title{Polymer chain in a quenched random medium:\\ slow dynamics and
   ergodicity breaking}
 
\author{ Gabriele Migliorini$^{1}$ Vakhtang G. Rostiashvili$^{1,2}$,and
   Thomas A.  Vilgis$^{1,2}$}\email{vilgis@mpip-mainz.mpg.de}
 \affiliation{$^1$ Max Planck Institute for Polymer Research\\
   10 Ackermannweg, 55128 Mainz, Germany.\\
   $^2$ Laboratoire Europ\'een Associ\'e, Institute Charles Sadron\\
   6 rue Boussingault, 67083 Strasbourg Cedex, France.}

\begin{abstract}
  The Langevin dynamics of a self - interacting chain embedded in a quenched
  random medium is investigated by making use of the generating functional
  method and one - loop (Hartree) approximation. We have shown how this
  intrinsic disorder causes different dynamical regimes. Namely, within the
  Rouse characteristic time interval the anomalous diffusion shows up. The
  corresponding subdiffusional dynamical exponents have been explicitly
  calculated and thoroughly discussed. For the larger time interval the
  disorder drives the center of mass of the chain to a trap or frozen state
  provided that the Harris parameter, $(\Delta/b^d) N^{2 - \nu d} \ge 1$,
  where $\Delta$ is a disorder strength, $b$ is a Kuhnian segment length, $N$
  is a chain length and $\nu$ is the Flory exponent. We have derived the
  general equation for the non - ergodicity function $f(p)$ which
  characterizes the amplitude of frozen Rouse modes with an index $p = 2\pi
  j/N$. The numerical solution of this equation has been implemented and shown
  that the different Rouse modes freeze up at the same critical disorder
  strength $\Delta_c \sim N^{-\gamma}$ where the exponent $\gamma \approx
  0.25$ and does not depend from the solvent quality.
\end{abstract}
 \maketitle

\section{Introduction}
The statical and dynamical properties of polymer chains in a random medium
bares still many open relevant questions in the current discussion on polymer
physics and disordered systems \cite{Baum}. This is due to the fact that this
model naturally occurs under the interpretation of many experiments: polymer
adsorption on heterogeneous surfaces \cite{Vilgis}, polymer diffusion in the
swollen polyvinylmethylate gels \cite{Lodge}, electrophoresis \cite{Viouy},
etc.

The theoretical investigation of the problem has been stimulated by the Monte
- Carlo simulation on a site - diluted lattice \cite{Baum1}.  These results
essentially extended to the dynamic properties (anomalous diffusion, chain
length, $N$, dependence of the diffusion coefficient $D$, etc.) in the number
of papers \cite{Baum2,Baum3,Muthu,Slater}. The authors of these papers have
found new dynamic laws where chain diffusion is slower than even reptation.
Such behavior emerges from the presence of ``entropic traps'', i.e. regions
which are relatively free from obstacles and hence entropically more
preferable. The diffusion is slowed down substantially by the presence of
narrow channels between traps, so that the chain is forced to squeeze through
them.

A large number of theoretical studies emphasizing the statical
\cite{Edwards,Cates,Machta,Machta1,Natter,Obukh,Harris,Kim,Step,Vilgis1} and
dynamical \cite{Machta,Mart,Step1,Ebert,Ebert1,Ebert2,Ebert3,Ebert4} aspects
of the problem. When discussing the static properties one should discriminate
between the annealed and quenched averaging. If the system is large enough and
the chain is still rather mobile, so that it experiences different disorder
environment (during the time of experiment) the problem is reduced to the
annealed one. In this case the only effect of disorder (which is characterized
by the second moment $\Delta$ of the quenched potential fluctuation) is the
reduction of the excluded volume parameter $v$, i.e. $v \to v - \Delta$
\cite{Natter,Step,Vilgis1,Ebert}. In the mean time it is well known
\cite{Natter,Step,Mart,Ebert,Ebert1} that the renormalization group (RG)
consideration for the discussed problem suffers from he lack of a stable fixed
point for the renormalized second moment $\Delta_{\rm R}$, so that the
perturbation theory can not be used. Nevertheless, it was argued
\cite{Ebert,Ebert1,Ebert2} that for the weak disorder in the thermodynamical
limit the quenched and annealed averaging lead to the same results. The
authors of ref.\cite{Harris,Kim} drew similar conclusions by studying a self -
avoiding chain on dilute lattices and making use the polymer - magnetic
analogy and $n \to 0$ trick \cite{Emery}.  On the other hand it has been shown
in ref. \cite{Machta,Machta1} that ``entropic traps'' caused by disorder
affect the equilibrium spatial distribution $Z_N$ of the polymer chain
strongly for $d < 4$. In this case the distribution is inhomogeneous, so that
typical and average values of $Z_N$ are different.  Under these conditions the
motion of the chain center of mass can be trapped.  The corresponding
diffusion coefficient, $D$, is scaled with the length of chain $N$ as $D \sim
\exp( - \Delta N^{\alpha})$, where the ``specific heat'' exponent $\alpha = 2
- \nu d$ and $\nu$ is the Flory exponent. It can be seen that for rather long
chains (remember  that $\alpha > 0$ at $d < 4$) the diffusion is effectively
suppressed and the whole system becomes nonergodic.

This is an indication of the multiple local minima configurations which could
manifest itself through the replica symmetry breaking (RSB) scheme
\cite{Dots,Dots1,Dots2}. It have been shown there that within the RG -
approach the conventional replica symmetric (RS) fixed point (which used to be
considered as providing new universal disorder - induced critical exponents)
are unstable with respect to an RSB - solution. In this case the structure of
the renormalized interactions develops strong RSB and the values of the
corresponding interaction parameters are getting large.

In this paper we use an alternative, dynamical, approach in order to treat
nonergodic regimes for the chain in a quenched random medium.  It is of
interest that the nonergodic, glassy - like regimes have been recently seen in
quasi - elastic neutron scattering experiments on dynamic properties of
flexible polymer chains filled with nano - particles (hydrophilic aerosil)
\cite{Gag}. The authors of \cite{Gag} have seen the enhancement of elastic
component of neutron scattering from chains whose dynamics is getting strongly
restricted and slowed down upon the growth of nano - particles fraction. To
the best of our knowledge the theoretical studies of these regimes is lacking.
We will use the Langevin dynamics and Martin - Siggia - Rose generating
functional method as well as the Hartree approximation \cite{Kinz} in order to
derive the equation of motion for our random model. We will show that at time
$t \to \infty$ this leads to the G\"otze - like equation \cite{Gotz} for the
non - ergodicity function, $f(p)$, which describes frozen Rouse - mode $p$ -
dependent states. The dynamic phase diagram in terms of Rouse mode indices and
disorder parameter $\Delta$ is numerically calculated and thoroughly
discussed.

\section{Preliminaries}
\subsection{Model}
The chain conformation is characterized by the d-dimensional vector - function
${\bf R}(s,t)$ of the time $t$ and $s$ (where $1 \le s \le N$), which labels
beads of the chain. Besides intrachain interactions the chain experiences a
quenched random external field, $V\{{\bf R}(s,t)\}$, so that the whole
Hamiltonian has the form
\begin{eqnarray}
 H=\frac{1}{2}\epsilon\sum_{s=0}^{N-1}\left[\nabla_{s}{\bf
     R}(s,t)\right]^{2} +  H_{\rm int}\left\{{\bf R}(s,t)\right\} + V\{{\bf
   R}(s,t)\} 
 \quad,
 \label{Hamilton}
\end{eqnarray}
where $\epsilon=dk_B T/b^{2}$ is the elastic modulus with the Kuhn segment
length $b$, $N$ is the length of chain and the finite difference
$\nabla_{s}{\bf R}_{j}(s,t) = {\bf R}_{j}(s + 1,t) - {\bf R}_{j}(s,t)$ . The
intra - chain Hamiltonian reads
\begin{eqnarray}  
  H_{\rm int}\left\{{\bf R}(s,t)\right\} &=&\frac{1}{2}
  \sum_{s=0}^{N-1}\sum_{s'=0}^{N-1}v({\bf R}(s,t) - {\bf R}(s',t))\nonumber\\
 &+&\frac{1}{3!} \sum_{s=0}^{N-1}\sum_{s'=0}^{N-1}\sum_{s''=0}^{N-1}w({\bf
    R}(s,t) - {\bf R}(s',t); {\bf R}(s',t) - {\bf R}(s'',t)) + \dots \quad,
 \label{Interaction}
\end{eqnarray}
where $v({\bf r})$ and $w({\bf r}_1, {\bf r}_2)$ are the second and third
virial coefficients correspondingly. The quenched random potential $V\{{\bf
  R}(s,t)\}$ is assumed to be Gaussian distributed, i.e.
\begin{eqnarray}
 \left < V({\bf r})V({\bf r'}) \right>  =   
 \Delta \delta^{(d)}( {\bf r} - {\bf r'}), 
 \label{Delta}
\end{eqnarray}
where dispersion $\Delta$ is one of the main control parameters of the
problem and has dimensions of  volume (in these units of measurement
$k_BT = 1$). This will become important below.
Eqs.(\ref{Hamilton}) - (\ref{Delta}) represent and specify our model.

\subsection{Annealed and quenched disorder}

Let us consider first the case when the characteristic times for chain's
configurations and chain's center of mass positions in a disordered medium are
of the same order of magnitude. Then in the course of an experiment the chain
experiences all possible quenched field realizations. This corresponds to the
annealed disorder and the corresponding free energy $F_{\rm anneal} = - \ln
\left< \Xi \right>_{V}$, where $\Xi$ is the partition function at a given
realization of $V({\bf r})$ and $\left< \dots \right>_{V}$ is the averaging
over the field $V({\bf r})$ distribution. It can be shown  that in this case
 \begin{eqnarray}
  \left< \Xi \right>_{V} = \Xi_{0} \{v - \Delta\}\quad,
 \label{Xi}
\end{eqnarray}
where $\Xi_{0} \{v\}$ is the partition function of the pure (i.e. without
disorder) system with the second virial coefficient $v$. As a result the only
effect of the disorder is the reduction of the second virial coefficient,
i.e., a reduction of the excluded volume. However, it may change the sign of
the effective second virial coefficient, and can cause therefore collapsed
states of the chain.

Nevertheless, in a medium with a strong disorder the chain is preferably
trapped in some regions where the depth of the quenched random potential
exceeds $k_B T$. In this case the chain is pinned down in some particular
place of a disordered medium and experiences only a local quenched field. This
corresponds to the quenched disorder and the relevant free energy $F_{\rm
  quench} = - \left< \ln \Xi \right>_{V}$.  For the site - diluted lattice
medium model it was argued by Machta \cite{Machta,Machta1} that while the size
of the chain is unaffected by the disorder (i.e. $R \sim b N^\nu$) the whole
spatial distribution of the chain is correlated with the disorder for $d < 4$.
The effect of the disorder shows up as an essential singularity in the $N$
dependence of the so - called typical value of the partition function,
$\Xi_{\rm typ} = \exp\left(- F_{\rm quench}\right)$, as well as the chain's
center of mass diffusion coefficient $D$.

The chain with a frozen center of mass explores only a local region of the
medium. As a result a simple estimation of the typical value of the partition
function reads \cite{Doussal}
\begin{eqnarray}
 \Xi_{\rm typ} \approx  \Xi_{0}\{v\}   \exp \Bigl\{ 
 \left(\frac{\Delta}{b^d}\right)^{1/2} N^{1
 - \nu d/2}\Bigr\}
 \label{typ}\quad,
\end{eqnarray}
where the disorder term comes from the dimensional analysis of the last term
in eq. (\ref{Hamilton}), which scale like $\Delta^{1/2} N/R^{d/2}$. 
The Flory free energy for a Gaussian chain is then simply given by
\begin{equation}
F_{\rm quench} =  \frac{R^2}{Nb^{2}} + 
\frac{Nb^2}{R^2} - \Delta^{1/2} \frac{N}{R^{d/2}} \quad.
\end{equation}
The first two terms corresponds to the stretching and compression entropy of
the Gaussian chain. The third term comes from the quenched disorder. It is
interesting for the discussion below, that the balance of the stretching term
and the disorder term yields a critical value for the disorder $\Delta_{\rm c}
\propto N^{-2+d/2}$,  whereas the comparison of the compression term and the
disorder term provides the localized radius $R \simeq b
(b^{d}/\Delta)^{1/(4-d)}$,  which agrees with various calculations of the
different type \cite{Baum,Vilgis1}. This result for Gaussian chains states
simply that the size of the chain is entirely determined by the disorder. The
chain size does no longer depend on the chain length $N$, i.e., it is
''localized'' in a typical volume $\Delta$.

The corresponding Flory free energy for excluded volume chains has the form
\cite{Doussal}
\begin{eqnarray}
F_{\rm quench} = - \ln \Xi_{\rm typ} \approx \frac{R^2}{Nb^{2}} +
\frac{Nb^2}{R^2} + v
\frac{N^2}{R^d} - \Delta^{1/2} \frac{N}{R^{d/2}} \quad,
\label{F_quench}
\end{eqnarray}
where the third term  corresponds  to  the  excluded volume interaction. Because for the long chain $N^{\alpha/2} \ll N^{\alpha}$, where
$\alpha = 2 - \nu d$, the excluded volume term dominates and the disorder
leaves the Flory exponent unchanged.

What can be expected from dynamic time scales? This can be estimated by
similar arguments.  Now we obtain for our random - field model a
straightforward estimation of the diffusion coefficient $D$. The
characteristic confinement time for the chain in a trap
 \begin{eqnarray}
 t^* \approx \tau_R \exp \Bigl\{ \sum_{s = 0}^{N - 1}  V\{{\bf R}(s,t)\}\Bigr\}
 \label{t*}\quad,
\end{eqnarray}
where $\tau_R$ is the maximal Rouse time. The confinement time $t^*$ has a
broad distribution according to fluctuations of barriers between traps. After
averaging over these fluctuations we have
\begin{eqnarray}
 \left< t^* \right>_{V}  &=&  \tau_R \exp \Bigl\{ \frac{\Delta}{2}
 \sum_{s = 0}^{N - 1}\sum_{s' = 0}^{N - 1} \: \delta \left({\bf R}(s,t) 
   - {\bf R}(s',t)\right)\Bigr\}\nonumber\\
 &\approx& \tau_R  \exp \Bigl\{ \frac{\Delta}{b^d}N^{2 - \nu d}\Bigr\}
 \label{t*_average}\quad,
\end{eqnarray}
where the second line is a direct \`a la Flory estimation of the $\delta$ -
function terms, provided that the size of of the chain $R \sim b N^\nu$.
 
By taking into account eq.(\ref{t*_average}) the estimation for $D$ reads
\begin{eqnarray}
 D \approx \frac{R^2}{\left< t^* \right>_{V}} = D_R \exp \Bigl\{ -
 \frac{\Delta}{b^d}N^{2 - \nu d}\Bigr\}
 \label{Diff_coeff}\quad,
\end{eqnarray}
which goes back to the Machta's result \cite{Machta,Machta1}. Since $\nu =
3/(d + 2)$ the ``specific - heat'' exponent $\alpha = 2 - \nu d = (4 - d)/(d +
2)$ and it can be seen  that at $d < 4$ and reasonably large $N$
\begin{eqnarray}
 \frac{\Delta}{b^d} \: N^{2 - \nu d} \gg 1
 \label{Harris}\quad,
\end{eqnarray}
so that the chain center of mass is effectively pinned down or frozen.

The inequality (\ref{Harris}) can be found independently as a paraphrase of
the well - known Harris criterion \cite{Dots} which express conditions when
the disorder completely dominates over the chain entropy. We have relegated
this consideration based on the $n$ - component field theory formulation
\cite{Kholod} to the Appendix A.

From the dynamical standpoint the chain's center of mass freezing leads to the
ergodicity breaking. It is a question of first importance to formulate a more
general dynamical approach which enable to study the slow dynamics and
ergodicity breaking not only for the center of mass but also for higher Rouse
modes. We will devote the rest of this paper to the developing of such
approach and show how in the long time limit the ergodicity breaking for
different Rouse modes appears.

 \section{Equation of motion for the time correlation function}
 \subsection{Langevin dynamics}
 
 In this section we give a general consideration of the Langevin dynamics of a
 polymer chain in the quenched random field.  The dynamics of the chain is
 described by the following Langevin equation
 \begin{eqnarray}
 \zeta_{0}\frac{\partial}{\partial
 t}R_{j}(s,t)&-&\epsilon\Delta_{s}R_{j}(s,t) +\frac{\delta}{\delta R_{j}(s,t)} H_{\rm
 int}\left\{{\bf R}(s,t)\right\} \nonumber\\
 &+& \frac{\delta}{\delta R_{j}(s,t)} V\{{\bf R}(s,t)\} = f_{j}(s,t)\quad,
 \label{R}
 \end{eqnarray}
 where $j$ labels Cartesian components, $\zeta_{0}$ is a bare friction coefficient and
 the second order finite difference $\Delta_{s}R_{j}(s,t) = R_{j}(s + 1,t) + R_{j}(s -
 1,t) - 2R_{j}(s,t)$.
 
The Langevin problem in question is getting much more convenient for the
theoretical investigation if we change to the MSR - generating functional
representation \cite{Rost}. The generating functional (GF) of our problem can
be written as
\begin{eqnarray}
 Z\left\{\cdots\right\}&=&\int DR_j(s,t)D{\hat
   R}_j(s,t)\nonumber\\
 &\times&\exp\left\{A_{\rm intra}\left[{\bf R}(s,t),\hat{\bf
   R}(s,t)\right] +  A_{\rm ext}\left[{\bf R}(s,t), \hat
   {\bf R}(s,t)\right]    \right\}\quad,
 \label{GF1}
 \end{eqnarray} 
 where the intra - chain action is given by
 \begin{eqnarray}
 A_{\rm intra}\left[{\bf R}(s,t),{\hat {\bf R}}(s,t)\right] &=& \sum_{s=0}^{N-1}\int
 dt\Bigg\{i{\hat R}_{j}(s,t)\left[\zeta_{0}\frac{\partial}{\partial
 t}R_{j}(s,t)-\epsilon\Delta_{s}R_{j}(s,t)\right]\nonumber\\
 &+&\frac{\delta}{\delta R_{j}(s,t)} H_{\rm
     int}\left\{R_{j}(s,t)\right\} + k_B T\zeta_{0}\left[i{\hat
       R}_{j}(s,t)\right]^{2}\Bigg\}
 \label{intraaction}
 \end{eqnarray}
 and the action related with the quenched random field reads 
 \begin{eqnarray}
 A_{\rm ext}\left[{\bf R}(s,t),{\hat {\bf R}}(s,t)\right] = \sum_{s=0}^{N-1}\int dt
 i{\hat R}_{j}(s,t)\frac{\delta}{\delta R_{j}(s,t)} V\{{\bf R}(s,t)\}\quad.
 \label{extaction}
 \end{eqnarray}
 The expressions (\ref{GF1}) - (\ref{extaction}) correspond to a given realization of
 the random field $V\{{\bf R}(s,t)\}$. Now we perform the averaging over all
 configurations of  $V\{{\bf R}(s,t)\}$ taking into account its Gaussian statistics
 (see eq.(\ref{Delta})). The resulting GF takes the following form
 \begin{eqnarray}
 \left<Z\left\{\cdots\right\}\right>_V  &=&\int DR_j(s,t)D{\hat
   R}_j(s,t)\exp\Bigg\{A_{\rm intra}\left[{\bf R}(s,t),\hat{\bf
 R}(s,t)\right]\nonumber\\ 
 &+& \Delta \sum_{s=0}^{N-1}\sum_{s'=0}^{N-1}\int dt dt' \int \frac{d^d k}{(2\pi)^d}
 k_j k_l \exp \left\{i {\bf k}[ {\bf R}(s,t) - {\bf R}(s',t')]\right\} i {\hat
 R}_j(s,t) i {\hat R}_l(s',t')\Bigg\}\quad.
 \label{GF2}
 \end{eqnarray}
 It can be seen from eq.(\ref{GF2}) that the averaging over the disorder leads
 to the non - Markovian (i.e. non - local in time) renormalization of the
 friction coefficient (which is coupled with $i {\hat R}_j(s,t) i {\hat
   R}_l(s',t')$). This causes actually dynamical slowing down and ergodicity
 breaking which we will discuss below.

 \subsection{Self - consistent Hartree approximation}
 
 In order to handle the functional integral (\ref{GF2}) we use the Hartree
 approximation. In this approximation the full MSR - action is replaced by the
 Gaussian one in such a way that all terms which include more than two fields
 ${\bf R}(s,t)$ and/or $\hat{\bf R}(s,t)$ are written in all possible ways as
 products of pairs of ${\bf R}(s,t)$ and/or $\hat{\bf R}(s,t)$ coupled to the
 self - consistent averages of the remaining fields. On the other hand in
 ref.\cite{Rost1} it was shown that the Hartree approximation is equivalent to
 taking into account Gaussian fluctuations around the saddle - point solution.
 The resulting Hartree action is a Gaussian functional with coefficients which
 could be represented in terms of correlation and response functions. The
 calculation of these coefficients is straightforward and details can be found
 in the Appendix B of ref.\cite{Rehk}. The second and third virial terms in
 $A_{\rm intra}\left[{\bf R}(s,t),\hat{\bf R}(s,t)\right]$ as well as the term
 which is responsible for the non - Markovian renormalization of the friction
 coefficient are treated in the same manner as in the ref. \cite{Rost2}.
 After collection of all these terms the final Hartree GF reads.
 \begin{eqnarray}
 \left<Z\{\cdots \}\right>_V &=& \int D{\bf R}D{\hat {\bf R}}\exp\Big\{A_{\rm
 intra}^{(0)}[{\bf R},{\hat {\bf
     R}}]\nonumber\\
 &+&\sum_{s=0}^{N-1}\sum_{s'=0}^{N-1}\int_{-\infty}^{\infty}dt\int_{-\infty}^{t}dt'\:i
 {\hat
   R}_{j}(s,t)R_{j}(s',t')\lambda(s,s';t,t')\nonumber\\
 &-&\sum_{s=0}^{N-1}\sum_{s'=0}^{N-1}\int_{-\infty}^{\infty}dt\int_{-\infty}^{t}dt'\:i
 {\hat
   R}_{j}(s,t)R_{j}(s,t)\lambda(s,s';t,t')\nonumber\\
 &+&\frac{1}{2}\sum_{s=0}^{N-1}\sum_{s'=0}^{N-1}
 \int_{-\infty}^{\infty}dt\int_{-\infty}^{\infty}dt'\:i{\hat
   R}_{j}(s,t)i{\hat R}_{j}(s',t')\chi(s,s';t,t')\Big\}\quad,
 \label{GF3}
 \end{eqnarray}
 where
 \begin{eqnarray}
 \lambda(s,s';t,t') &=&
 \frac{\Delta}{d}G(s,s';t,t')\int\frac{d^{d}k}{(2\pi)^{d}}k^{4}F({\bf
 k};s,s';t,t')\nonumber\\
 &+&\int\frac{d^{d}k}{(2\pi)^{d}} \: k^{2} v({\bf k}) F_{\rm st}({\bf
 k};s,s')\nonumber\\
 &+&\sum_{s''=1}^{N}\int\frac{d^{d}kd^{d}q }{(2\pi)^{2d}} \: k^2 w({\bf k},{\bf
   q})F_{\rm st}({\bf q};s',s'')F_{\rm st}({\bf k};s,s')
 \label{Lambda}
 \end{eqnarray}
 and
 \begin{eqnarray}
 \chi(s,s';t,t')= \Delta \int\frac{d^{d}k}{(2\pi)^{d}}\: k^{2}F_{\rm st} ({\bf
 k};s,s')
 \label{Chi}.
 \end{eqnarray}
 In eqs.(\ref{Lambda}) - (\ref{Chi}) the response function 
 \begin{eqnarray}
 G(s,s';t,t')=\left<i{\hat {\bf R}}(s',t'){\bf R}(s,t)\right>
 \label{G}
 \end{eqnarray}
 and the chain density correlator
 \begin{eqnarray}
 F({\bf k};s,s';t,t')=\exp\left\{-\frac{k^{2}}{d}Q(s,s';t,t')\right\}
 \label{F}
 \end{eqnarray}
 with
 \begin{eqnarray}
 Q(s,s';t,t')\equiv\left<{\bf R}(s,t){\bf R}(s,t)\right>-\left<{\bf
     R}(s,t){\bf R}(s',t')\right> \quad,
 \label{Q}
 \end{eqnarray}
 whereas $F_{\rm st}({\bf k};s,s')$ stands for the static limit of (\ref{F}). 
 The pointed brackets denote the self - consistent averaging with the Hartree GF which
 is given by eq.(\ref{GF3}).
 
 In general one should consider fluctuation dissipation theorem (FDT)
 violation which is well known in the context of glass transition phenomenon
 \cite{Bouch}. In our present consideration we are mainly interested in the
 freezing conditions as well as the anomalous diffusion at the relatively
 short times. This enables us to assume that the FDT and the time
 translational invariance (TTI) are  valid, then
 \begin{eqnarray}
 G(s,s';t-t') = (k_B T)^{-1}\frac{\partial}{\partial t'} \:\: Q(s,s';t-t')\:\:\:\:{\rm
 at}\:\:\: t>t'
 \label{FDT1}
 \end{eqnarray}
 By employing eq.(\ref{FDT1}) in eqs.(\ref{GF3}) - (\ref{Chi}) and after
 integration by parts with respect to time argument $t'$, we obtain the
 following Hartree GF:
 \begin{eqnarray}
 \left<Z\{\cdots \}\right>_V &=& \int D{\bf R}D{\hat {\bf
 R}}\exp\Big\{\sum_{s,s'=0}^{N-1}\int_{-\infty}^{\infty}dt\int_{-\infty}^{t}dt'\:i{\hat{R}}_{j}(s,t)\left[ \zeta_0 \delta(t - t') + \theta(t -
   t')\Gamma(s,s';t,t')\right]\frac{\partial}{\partial t}R_{j}(s',t')\nonumber\\
 &-&\sum_{s,s'=0}^{N-1}\int_{-\infty}^{\infty}dt\int_{-\infty}^{t}dt'\:i{\hat
   R}_{j}(s,t)\:\Omega(s,s')\:R_{j}(s',t)\nonumber\\
 &+& \sum_{s,s'=0}^{N-1}
\int_{-\infty}^{\infty}dt\int_{-\infty}^{\infty}dt'\:i \hat{R}_{j}(s,t)
\left[ \zeta_0 \delta(t - t') + \theta(t - t')\Gamma(s,s';t,t')\right] 
i{\hat R}_{j}(s',t')\Big\}\quad,
 \label{GF4}
\end{eqnarray}
 where the memory function
\begin{eqnarray}
 \Gamma(s,s';t,t')= \Delta  \int\frac{d^{d}k}{(2\pi)^{d}} \:k^{2}F({\bf k};s,s';t,t')
 \label{Memory}
\end{eqnarray}
and the effective elastic susceptibility
\begin{eqnarray}
 \Omega(s,s') &=& \epsilon\delta_{\rm ss'}\Delta_{\rm s} -
 \int\frac{d^{d}k}{(2\pi)^{d}} \: k^{2} \left(v({\bf
   k}) - \Delta \right) \left[F_{\rm st} ({\bf k};s,s') - \delta_{\rm
     ss'}\sum_{s''=0}^{N-1} F_{\rm st}({\bf k};s,s'')\right]\nonumber\\
 &-&\frac{1}{2}\sum_{s''=0}^{N-1}\int\frac{d^{d}kd^{d}q}{(2\pi)^{2d}} \:
 k^2 w({\bf k}, {\bf q})\nonumber\\
 &\times&\left[F_{\rm st}({\bf k};s,s')F_{\rm st}({\bf q};s'',s') - \delta_{\rm
 ss'}\sum_{s'''=0}^{N-1} F_{\rm st}({\bf k};s,s''')F_{\rm st}({\bf
 q};s''',s'')\right]\quad.
 \label{Omega}
\end{eqnarray}
In eqs.(\ref{GF4}) - (\ref{Omega}) we use for simplicity the units where $k_B
T = 1$, so that the disorder parameter $\Delta$ has the dimensionality of
volume. The memory function (\ref{Memory}) is responsible for the non -
Markovian renormalization of the Stokes friction coefficient $\zeta_0$ which
arises from interaction with the quenched field $V({\bf k})$. The effective
elastic susceptibility (\ref{Omega}) takes into account all non - dissipative
(reactive) forces in the system: local spring interaction, renormalization of
the second virial coefficient due to the random field $V({\bf k})$ as well as
third virial term.

\subsection{Equation of motion}

The equation of motion for the correlation function
\begin{eqnarray}
 C(s,s';t,t') = \left<{\bf R}(s,t){\bf R}(s',t')\right>
 \label{TD-correlator}
\end{eqnarray}
can readily be obtained from GF (\ref{GF4}). The result at $t > t'$ reads
\begin{eqnarray}    
 \zeta_0 \frac{\partial}{\partial t}C(s,s';t,t') &-& \sum_{m = 1}^{N} \:
 \Omega (s,m;t) C(m,s';t,t')\nonumber\\
  &+&  \sum_{m = 0}^{N-1}\int_{t'}^{t} \:
 \Gamma(s,m;t,\tau) \frac{\partial}{\partial \tau}C(m,s';\tau,t') d \tau = 0 \quad.
\label{EqMotion1}
\end{eqnarray}
 It is convenient to make the Rouse transformation \cite{Doi}
 \begin{eqnarray}
 C(p,t) = \frac{1}{N}\sum_{s = 0}^{N-1} C(s,t)
 \exp (is p)
\label{Fourier1}
\end{eqnarray}
 and
\begin{eqnarray}
 C(s,t) = \sum_{p = 0}^{2\pi} C(p,t)
 \exp (- i s p) \quad,
 \label{Fourier2}
\end{eqnarray}
where $p = 2\pi j / N  \quad  (j = 0,1, \dots N - 1)$, i.e. we have used for
simplicity the cyclic boundary conditions.  After this transformation the
eq.(\ref{EqMotion1}) is simplified and takes the form
\begin{eqnarray}
 \zeta_0 \frac{\partial}{\partial t}\:C(p;t) + N \int_{0}^{t}\Gamma(p, t -
 t')\:\:\frac{\partial}{\partial t'}\:\:C(p;t')dt' + \Omega (p, t) \:C(p;t) = 0
 \label{EqMotion2}
\end{eqnarray}
where
\begin{eqnarray}
 N \Gamma (p, t) =  \Delta \:\: \frac{d^{\frac{d}{2} + 2}}{2^{d+1} \pi^{d/2}}
 \:\sum_{n = 0}^{N-1} \:\:\frac{\cos(p s)}{\left[Q(n,t)\right]^{\frac{d}{2} + 1}}
 \label{Memory1}
\end{eqnarray}
and
\begin{eqnarray}
 \Omega(p) &=& \frac{2d}{b^2} (1 - \cos p) - \left(v - \Delta\right)\:
 \frac{d^{\frac{d}{2} 
     + 2}}{2^{d+1}
   (\pi)^{\frac{d}{2}}}\sum_{n = 0}^{N-1}
 \: \frac{1 - \cos(p n)}{\left[Q_{\rm st}(n)\right]^{\frac{d +
       2}{2}}}\nonumber\\
 &-& w \:\frac{d^{d + 2}}{4^{d+1}
   (\pi)^{d}}\sum_{n = 0}^{N-1}\:\:\sum_{m =
   0}^{N - n - 1}\: \frac{1 - \cos(p n)}{\left[Q_{\rm st}(n)\right]^{\frac{d +
       2}{2}}\left[Q_{\rm st}(n)\right]^{\frac{d}{2}}}\quad.
 \label{OmegaFourier1}
\end{eqnarray}
 In eqs. (\ref{Memory1}) and (\ref{OmegaFourier1}) the time dependent mean - square
 distance
\begin{eqnarray}
 Q (s, t) &=& \frac{1}{2} \left<\left[{\bf R}(s, t) - {\bf R}(0, 0)\right]^2
 \right>\nonumber\\
 &=& \sum_{p = 0}^{2\pi} \left[ C_{\rm st} (p) - \cos(p s) C(p, t)\right]
 \label{Q1}
\end{eqnarray}
 as well as its static limit
 \begin{eqnarray}
 Q_{\rm st}(s) = \sum_{p = 0}^{2 \pi} \left[ 1 - \cos(p s)\right] C_{\rm st}
 (p)\label{Q2} 
\end{eqnarray}
make the whole equation of motion for $C(p, t)$ self - consistently closed.
In the course of derivation of eqs.(\ref{EqMotion2}) - (\ref{OmegaFourier1})
we have took into account that the segment - segment interaction is short -
ranged, i.e.  $v({\bf k}) \approx v$ and $w({\bf k}, {\bf q})\approx w$; we
have used also the Rouse transformation of the chain density correlator, i.e.
\begin{eqnarray}
 F({\bf k} ; p ;t) = \frac{1}{N} \sum_{n = 0}^{N-1} \cos (p
 n)\exp\left\{ - \frac{k^2}{d} Q(n, t)\right\}\quad.
 \label{F-correlator}
\end{eqnarray}
The static limit (i.e. $t \to 0$) is evident from eq. (\ref{EqMotion2})
provided that the initial condition \cite{Rost}
 \begin{eqnarray}
 \zeta_0 \left(\frac{\partial}{\partial t}\:C(p;t)\right)_{t \to 0^{+}} = \zeta_0 G(p,
 t \to 0^{+}) = - \frac{d}{N}
 \label{Initial}
 \end{eqnarray}
 is taken into account. Then the static equation becomes
\begin{eqnarray}
 \left[N C_{\rm st}(p)\right]^{-1} &=& \frac{2}{b^{2}} (1 - \cos p) - \left(v - \Delta
 \right)\:\:
 \frac{d^{\frac{d}{2}+1}}{2^{d+1}
   (\pi)^{\frac{d}{2}}}\sum_{n = 0}^{N-1}
 \: \frac{1 - \cos(p n)}{\left[Q_{\rm st}(n)\right]^{\frac{d +
       2}{2}}}\nonumber\\
 &-& w \:\:\frac{d^{d+1}}{4^{d+1} 
   (\pi)^{d}}\sum_{n = 0}^{N-1}\:\:\sum_{m =
   0}^{N - n -1}\: \frac{1 - \cos(p n)}{\left[Q_{\rm st}(n)\right]^{\frac{d +
       2}{2}}\left[Q_{\rm st}(n)\right]^{\frac{d}{2}}}\quad.
 \label{Static}
\end{eqnarray}
It is of interest that this equation is similar, with an accuracy of
prefactors and shifting $v \to v - \Delta$, to the variational equation, which
we have derived in ref.\cite{Migl}. In the static limit this shifting is an
only consequence of the random field $V({\bf r})$ effect.

\section{Dynamic behavior of the chain}
 
We are now in position to launch a more elaborate investigation of the chain
dynamic behavior which is based on eqs.(\ref{EqMotion2}) - (\ref{Q1}).  There
are at least two subjects which can be studied: i) the anomalous diffusion on
the  interval between a microscopic characteristic time  $\tau_{\rm
  d}$ (see below)  and the longest internal
relaxation time $\tau_R$ \cite{Baum,Ebert}; ii) Rouse modes dynamical freezing
at $t \to \infty$.

\subsection{Anomalous diffusion}  

The presence of the quenched random field restricts the motion of the chain
already at the time interval,
\begin{eqnarray}
 \tau_{\rm d} < t < \tau_0 N^{1 +2\nu}  
 \label{interval}
\end{eqnarray}
where $\tau_{\rm d}$ is a crossover time when the disorder starts to show up
(the value of $\tau_{\rm d}$ will be discussed below) and $\tau_0 N^{1 +2\nu}$
is the maximal Rouse time \cite{Doi} with the Flory exponent $\nu$.  This
restriction manifest itself through the subdiffusional regimes (anomalous
diffusion) which have been seen first by Monte Carlo (MC) simulation
\cite{Muthu}.

 Let us start from the general solution of eq.(\ref{EqMotion2}). For the Laplace
 correlator 
 \begin{eqnarray}
 C(p, z) = \int_{0}^{\infty} dt C(p, t) \exp ( -z t )
 \label{Laplace}
 \end{eqnarray}
 this solution reads \cite{Rost}
 \begin{eqnarray}
 C(p, z) = \frac{C_{\rm st}}{ z + \frac{\Omega (p)}{\zeta_0 + N \Gamma (p, z)}}
 \label{Fraction}
 \end{eqnarray}
 The calculation of $\Gamma (p, z)$ is based on eq.(\ref{Memory1}),
 where the time dependent mean - square distance $Q(s, t)$ at the time
 interval (\ref{interval}) is approximated by
 \begin{eqnarray}
 Q(s, t) = b^2 \left(\frac{t}{\tau_0}\right)^{2\theta} + Q_{\rm st}(s)
 \label{distance}
 \end{eqnarray}
 where $\theta = \nu/(1 + 2\nu)$ and $Q_{\rm st}(s) = b^2 s^{2\nu}$.
 This form can be justified by implementing simple scaling arguments
 for a pure (i.e. without disorder) model \cite{Rost2}. The substitution of
 eq.(\ref{distance}) in eq.(\ref{Memory1}) leads to the following
 result:
 \begin{eqnarray}
 N\Gamma(p \to 0, t) = A\: \left(\frac{\Delta}{b^{d + 2}}\right)
 \left(\frac{\tau_0}{t}\right)^{\beta} 
 \label{Memory2}
 \end{eqnarray}
 where the constant
 \begin{eqnarray}
 A = \frac{d^{\frac{d}{2} + 2}
   {\tilde\Gamma}\left(\frac{1}{2\nu}\right)
   {\tilde\Gamma}\left(\frac{d}{2} - \frac{1}{2\nu} + 1\right)}{2^{d +
     2} \pi^{\frac{d}{2}} \nu {\tilde\Gamma}\left(\frac{d}{2} + 1\right)}
 \label{constant}
 \end{eqnarray}
 with ${\tilde\Gamma}\left(x\right)$ stands for the gamma function and
 the exponent
 \begin{eqnarray}
 \beta &=& \theta \left(d + 2 - \frac{1}{\nu}\right)\nonumber\\
 &=& 1 - \frac{\alpha}{2\nu + 1} < 1 \quad,
 \label{Exponent}
 \end{eqnarray}
 where $\alpha = 2 - \nu d$ is the ``specific - heat'' exponent.

 The Laplace transformation of (\ref{Memory2}) at $\tau_0 z \ll 1$
 reads
 \begin{eqnarray}
 N\Gamma(p \to 0, t) = A\: \left(\frac{\Delta}{b^{d + 2}}\right)
 \tau_0^\beta \left(\frac{1}{z}\right)^{1 - \beta}
 \label{Memory3}
 \end{eqnarray}
 In the case when the memory term is in excess of the bare friction
 coefficient, i.e. at $t > \tau_{\rm d}$, we can use eq.(\ref{Memory3}) in 
 eq.(\ref{Fraction}) which after inverse Laplace transformation  can be
 put in the form (see   the ref.\cite{Rost})
 \begin{eqnarray}
 C(p,t)=C_{\rm st}(p)\sum_{k=0}^{\infty}\frac{\left[-\left(\frac{b^{d + 
           2} \Omega(p)}{\Delta
 A}\right)\left(\frac{t}{\tau_{0}}\right)^{\beta}\right]^{k}}{{\tilde\Gamma}(k\beta
 +1)}
 \label{Stretched}
 \end{eqnarray}
 Center of mass mean square displacement
 \begin{eqnarray}
 Q_{\rm c.m.}(t) &=& \frac{1}{2} \left< \left[ {\bf R}_{\rm c.m.}(t)  -
     {\bf R}_{\rm c.m.}(0)\right]^2 \right>\nonumber\\
 &=& \lim_{p \to 0} \left\{C_{\rm st}(p) - C(p, t)\right\}
 \label{Center_of_mass}
 \end{eqnarray}
 The substitution of eq.(\ref{Stretched}) in eq.(\ref{Center_of_mass})
 results in the leading term of the anomalous diffusion, i.e.
\begin{eqnarray}
 Q_{\rm c.m.}(t) = \frac{{\cal D}_0}{N}\left(\frac{t}{\tau_0}\right)^{\beta}
 \label{Center_of_mass1} \quad,
 \end{eqnarray}
 where
\begin{eqnarray}
 {\cal D}_0 = \frac{b^{b + 2}}{\Delta A}
 \label{D_0}
\end{eqnarray}
In the course of derivation of eq.(\ref{Center_of_mass1}) we have used the
static equation (\ref{Static}), i.e. $C_{\rm st}(p) \Omega(p) = 1/N$.

It is easy now to estimate the crossover time $\tau_{\rm d}$ after which the
disorder starts to effect the diffusion (see eq.(\ref{interval})). The
condition for that, $\zeta_0 = \int_{0}^{\tau_{\rm d}}\: dt \:\Gamma(p \to 0,
t)$ can be recast in a form
\begin{eqnarray}
 \tau_{\rm d}  = \left(\frac{b^{d + 2}\: \zeta_0}{\Delta\: A
     \:\tau_0^\beta}\right)^{\frac{1}{1 - \beta}}
 \label{crossover}
\end{eqnarray}
One can see that the anomalous diffusion exponent $\beta$ does not depend from
the strength of disorder, whereas the prefactor ${\cal D}_0$ decreases with
increasing $\Delta$. For a chain in the good solvent  $\nu = 3/(d + 2)$ and at
$d = 3$ the exponent $\beta_{\rm SAW} = 0.9$. For a Gaussian chain $\nu = 1/2$
and $\beta_{\rm Gauss} = 0.75$, i.e. the subdiffusional exponent has the same
value as in a polymer melt \cite{Rost}. Finally in the case of the globule
state $\nu = 1/3$ and $\beta_{\rm Globule} = 0.4$, i.e. the globule anomalous
diffusion is suppressed down by the disorder at most.

\subsection{Center of mass freezing}

Let us consider now the large time center of mass diffusion. In this case the
characteristic time interval
\begin{eqnarray}
 t \gg \tau_0 N^{1 + 2\nu}
 \label{interval1}
\end{eqnarray}
 and internal Rouse modes are already relaxed. For this time regime the 
 reasonable approximation for $Q(n, t)$ has the following form (compare 
 with eq.(\ref{distance})):
 \begin{eqnarray}
 Q(s, t) = d D t  + Q_{\rm st}(s) \quad,
 \label{distance1}
\end{eqnarray}
 where $D$ is the full (not bare) diffusion coefficient, which is
 renormalized by the effect of disorder and should be find self -
 consistently. On the other hand the equation for the zero - mode
 diffusion coefficient has the form \cite{Rost,Rehk,Hess}
\begin{eqnarray}
 D =  \frac{1}{N \left[\zeta_0 + N \int_{0}^{\infty} dt \Gamma(p = 0,t)\right]}
\label{Diffusion}
\end{eqnarray}
(we recall that in our units of measurement $k_B T = 1$). Eq.
(\ref{Diffusion}) enables to find $D$ self - consistently. By making use
eqs.(\ref{Memory1}) and (\ref{distance1}) in eq.(\ref{Diffusion}) we obtain
the following result for the center of mass diffusion coefficient
\begin{eqnarray}
 D = D_{\rm R} \left(1 - \Delta \: {\cal F}_{\rm N}\right) \quad,
\label{Diffusion1}
 \end{eqnarray}
 where $D_{\rm R} = (\zeta_0 N)^{-1}$ is the Rouse diffusion
 coefficient and
 \begin{eqnarray}
  {\cal F}_{\rm N} = \frac{d^{d/2}}{2^d \pi^{d/2}} \: N \: \sum_{s =
    0}^{N - 1} \: \frac{1}{\left[ Q_{\rm st}(s)\right]^{d/2}}\quad.
 \label{cal_F} 
\end{eqnarray}
It can be  seen that at $\Delta {\cal F}_{\rm N} \ge 1$ the center of mass
diffusion is frozen and the system becomes nonergodic. The relevance of this
result is twofold. First  this is a particular case of the so called
$A$ - type dynamical phase transition which has been extensively discussed
\cite{Gotz1} in the context of the mode coupling theory. On the other hand if
we substitute $ Q_{\rm st}(s)$ in eq.(\ref{cal_F}) with its most
representative term $Q_{\rm st} \approx b^2 N^{2\nu}$ we will find $D \approx
D_{\rm R} \left[1 - {\rm const} (\Delta/b^d) N^{2 - \nu d}\right]$. As a
result we turn back to the Machta's formula or, more exactly, to its expansion
up to the first order with respect to $(\Delta/b^2) N^{2 - \nu d}$. This
actually means that eq. (\ref{Diffusion1}) overestimate the freezing and one
should rather treat eq. (\ref{Diffusion1}) as a crossover criterion for the
weak ergodicity breaking transition in the ref. \cite{Bouch1} sense.

\subsection{Rouse modes freezing and a two mode  toy model}

Now we study the freezing or the ergodicity breaking of Rouse modes with $p
\ne 0$. This phenomenon mathematically manifests itself as a bifurcation with
respect to the non - ergodicity function which, in its turn, is a long time
limit of the corresponding correlator \cite{Gotz,Gotz1}. Let us define such
persistent part of the normalized correlator (i.e. non - ergodicity function)
as the long time limit
\begin{eqnarray}
 f(p) = \lim_{t \to \infty} \:  \frac{C(p, t)}{C_{\rm st}(p)} \quad.
 \label{non_ergodicity}
\end{eqnarray}
The equation for $f(p)$ can be easily obtained by taking the limit $t \to
\infty$ in eq.(\ref{EqMotion2}). The result reads
 \begin{eqnarray}
 \frac{f(p)}{1 - f(p)} = \Delta \frac{d^{\frac{d}{2} + 1}}{2^{d + 1}
   \pi^{\frac{d}{2}}} \:N C_{\rm st}(p)\: \sum_{s = 0}^{N - 1} \:
 \frac{\cos(ps)}{\left[L(s)\right]^{\frac{d}{2} + 1}} \quad,
 \label{Gotze_eq}
\end{eqnarray}
where
\begin{eqnarray}
 L(s) = \sum_{p = 2\pi/N}^{2\pi} \:  C_{\rm st}(p)\left[ 1 - \cos (qs) 
   f(q)\right] \quad.
\label{LL}
\end{eqnarray}
 The eq.(\ref{Gotze_eq}) is a self - consistent equation for the non -
 ergodicity function $f(p)$.
 In the vicinity of the bifurcation point the non - ergodicity function
 $f(p)$  is small and we can expand the r.h.s. of eq. (\ref{Gotze_eq})
 with respect to $f(p)$. It is shown in the Appendix B  that because of 
 orthogonality the zero - order term of this expansion vanishes and we
 arrive at the so - called $F_{\rm 12}$ - model according G\"otze's
 nomenclature \cite{Gotz}. The extensive numerical analysis of the full 
 eq.(\ref{Gotze_eq}) which is given in the next section reveals that
 the bifurcation of $f(p)$ is continuous or of $A$ - type. 

To gain a better insight into the Rouse modes freezing mechanism let
us consider first a  simplified version of eqs.(\ref{F_12}). This is
a toy model which is based on the truncation of the
full hierarchical eqs.(\ref{F_12}) on the level of  two longest modes, $j = 1$ 
and $j = 2$. In this case the asymptotic form $N C_{\rm st} \approx
p^{-1-2\nu}$ where $p \approx 2\pi j/N \ll 1$ can be used in order to
calculate the coefficients in eqs.(\ref{F_12}). As a result the toy
model equations for $f(1) \equiv f$ and $f(2) \equiv g$ can be recast
in the following form
\begin{eqnarray} 
\frac{f}{1 - f} &=& \Delta_1 f + \epsilon_1 f g \nonumber\\
\frac{g}{1 - g} &=& \Delta_2 g + \epsilon_2 f^2 \quad,
\label{toy}
\end{eqnarray}
where the coefficients 
\begin{eqnarray} 
\Delta_1 &=& \Delta N^{2 - \nu d} \quad\quad\quad\quad \epsilon_1  =
\frac{\Delta N^{2 - \nu d}}{2^{2\nu}}\nonumber\\
\Delta_2 &=& \frac{\Delta N^{2 - \nu d}}{2^{2 +
    4\nu}}\quad\quad\quad\quad\epsilon_2  = 2^{1 + 2\nu} \Delta N^{2 - \nu d}\quad.
\label{coefficients}
\end{eqnarray}
It should be mentioned that the important $f^2$ - term in the second
eq. (\ref{toy}) comes from the truncation of the last sum in r.h.s. of 
eq.(\ref{F_12}).

It is readily seen that in the vicinity of the critical point,
$\Delta_1^c = 1$, the coefficient $\Delta_1 = 1 + \sigma$, where
$\sigma \ll 1$ and mode amplitudes have  the following forms: $f
\approx \sigma f_+$ and $g \approx \sigma^2 g_+$, where $f_+$ and
$g_+$ are some constants. The substitution of these forms in
eq.(\ref{toy}) leads to the solution
\begin{eqnarray}
f(\sigma) &=& \sigma \nonumber\\
g(\sigma) &=& \sigma^2 \frac{\epsilon_2^c}{1 - \Delta_2^c} \quad,
\end{eqnarray}
where it is important that $\Delta_2^c < 1$.

As a result the trivial solution, $f = g = 0$, bifurcates at the
critical point $\Delta_1 = 1$, so that the  f-mode goes  linearly and
the g-mode -  quadratically with respect to $\sigma$. It is obvious  that 
close to the critical point (i.e. $\sigma \ll 1$) there is no effect
of the g-mode on the f - mode. On the other side g - mode bifurcates
only as a result of f - mode bifurcation. In this respect one can say
that the Rouse mode freezing follows the  ``host - slave'' scenario. In
Sec.V we will show that this scenario holds true for the whole
numerical solution.

\section{Numerical analysis}

In this section we present the numerical solution of eqs.(\ref{Gotze_eq}) -
(\ref{LL}) in the full range of the Rouse mode index $j$  values and for
increasing values of the disorder strength $\Delta$ which here acts as a
control parameter.  As it usually is in the mode coupling theory \cite{Gotz}
the full information about the static correlator, $C_{\rm st}(p)$, is a
necessary prerequisite for the non - ergodicity equation study. In this
respect, considering a chain of given length, we have numerically solved the
static equation (\ref{Static}) for $C_{\rm st}(p)$, where the virial
coefficients and the disorder strength $\Delta$ are given. By making use of
the Fast Fourier Algorithm we have implemented the bisection procedure between
two trial profiles of $C_{\rm st}(p)$ until the convergence to the final
solution is achieved. This method has been recently used in the different
context \cite{Migl} where it has enabled to consider chains of length up to $N
= 2^{8}$. After that we use $C_{\rm st}(p)$ as a static input for the non -
ergodicity equation (\ref{Gotze_eq}) - (\ref{LL}). This equation is solved for
the chain length $N = 128$ in much the same way as it is described above for
the static calculation.

\subsection{Bifurcation diagram}

We have found that for the small values of the disorder strength $\Delta$ the
only solution of eq.(\ref{Gotze_eq}) turn out to be the trivial one, i.e.
$f(p) = 0$. As the disorder strength increases above a critical value
$\Delta_{\rm c}$, we observe that the first and all other modes become frozen
simultaneously, i.e. they are characterized by a non vanishing value of the
non-ergodicity function $f(p)$ at the same $\Delta_{\rm c}$. The resulting
bifurcation phase diagram is shown in Fig.{\ref{Diagram}.
\begin{figure}[ht]
\begin{center}
\begin{minipage}{11cm}

    \centerline{\includegraphics[width=8cm]{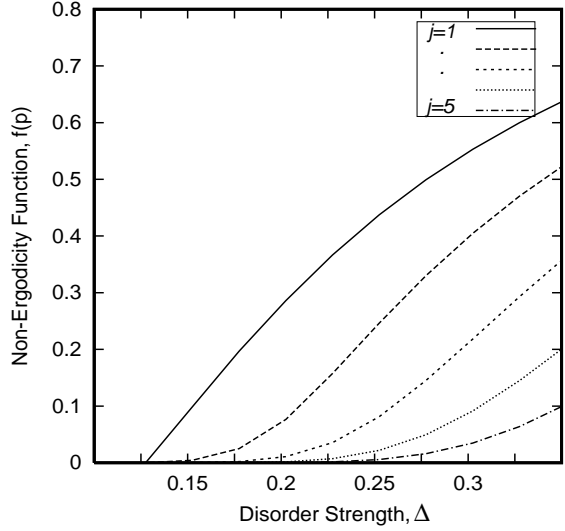}} \vspace*{10pt}

      \caption{\label{Diagram}
        Bifurcation diagram for the non - ergodicity function according to
        equation (\ref{Gotze_eq}). The calculation refers to a chain of
        $N=128$ monomers and the bare second virial coefficient $v = 0$. We
        show only the smaller mode index values, i.e. $j=1,..,5$.  The
        freezing of the modes appears above a critical value of $ \Delta_{\rm
          cr} \approx 0.13$. }
\end{minipage}
\end{center}
\end{figure}
As may be seen from Fig.\ref{Diagram} all modes bifurcate continuously
(A-type) , but bifurcations of higher modes ($j=2,3, \dots$) are more smooth
compare to the first mode bifurcation. This is qualitatively consistent with
the result of the toy model analysis from the Sec.IV C. Moreover, one can see
that the higher the Rouse mode index the more smooth is the bifurcation.  This
creates some numerical difficulties in precise location of the higher modes
critical point $\Delta_{\rm c}$. As soon as the accuracy of a non - vanishing
value of $f(p)$ (or the resolution) is not high enough the bifurcation diagram
looks as if there were a subsequent mode freezing (see Fig.\ref{3D_Diagram}).
We will show in the Sec.V B by the careful analysis of the finite resolution
problem that this is an illusory effect and all modes do freeze at the same critical point.
\begin{figure}[ht]
 \begin{center}
   \begin{minipage}{11cm}
     \centerline{\includegraphics[width=8cm]{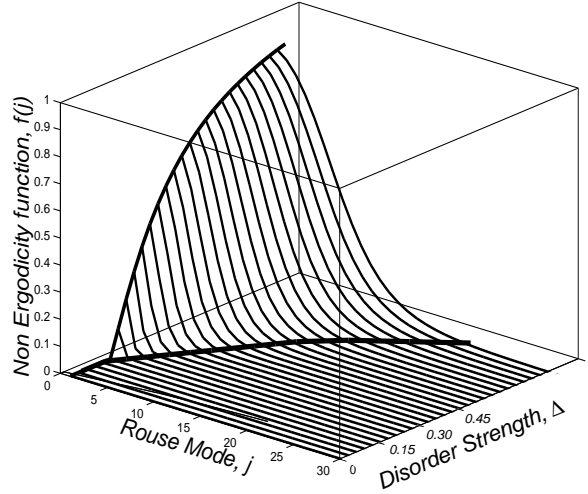}}
     \vspace*{10pt}
     \caption { \label{3D_Diagram}
       3D - Bifurcation diagram. The illusory Rouse mode successive freezing
       is a result of a finite resolution. In this case the accuracy of a non
       - zero value of $f(p)$ or the resolution $h = 10^{-4}$. }
\end{minipage}
\end{center}

\end{figure}

\begin{figure}[ht]
  \begin{center}
  \begin{minipage}{8cm}
    \centerline{\includegraphics[width=7cm,angle=270]{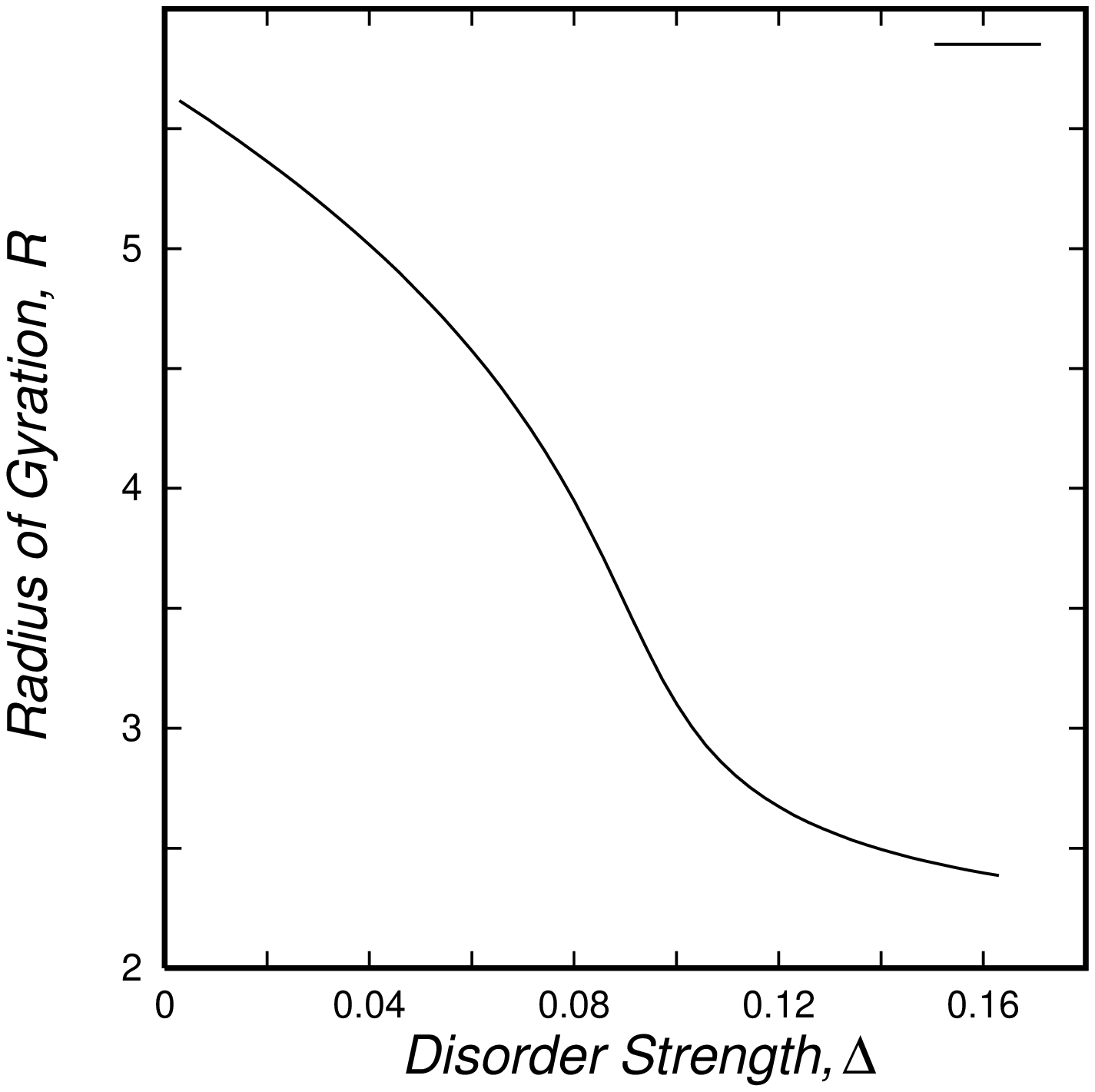}} \vspace*{10pt}
      \caption{ \label{Gyration_Rad} Radius of gyration as a function of disorder strength $\Delta$ for the chain length $N = 128$ and $v = 0$.
         }

  \end{minipage}
 \end{center}
\end{figure}

The critical value $\Delta_{\rm c}  \approx 0.13$  should be
correlated with the radius - of -gyration diagram in
Fig.\ref{Gyration_Rad}. It can be clearly seen that at disorder strength
comparable with this value the system is approaching the radius of
gyration which corresponds to the globular phase \cite{Migl}. We
recall that the static input information which is embraced by
eq.(\ref{Gotze_eq}) is determined  by the effective virial coefficient $v_{\rm eff} = v - \Delta$, where $v$ is a bare second virial coefficient. That is why the bifurcation diagram in Fig.\ref{Diagram} corresponds to the Rouse mode freezing in the globule phase.

\begin{figure}[ht]
\begin{center}
  \begin{minipage}{9cm}
    \centerline{\includegraphics[width=7cm,angle=270]{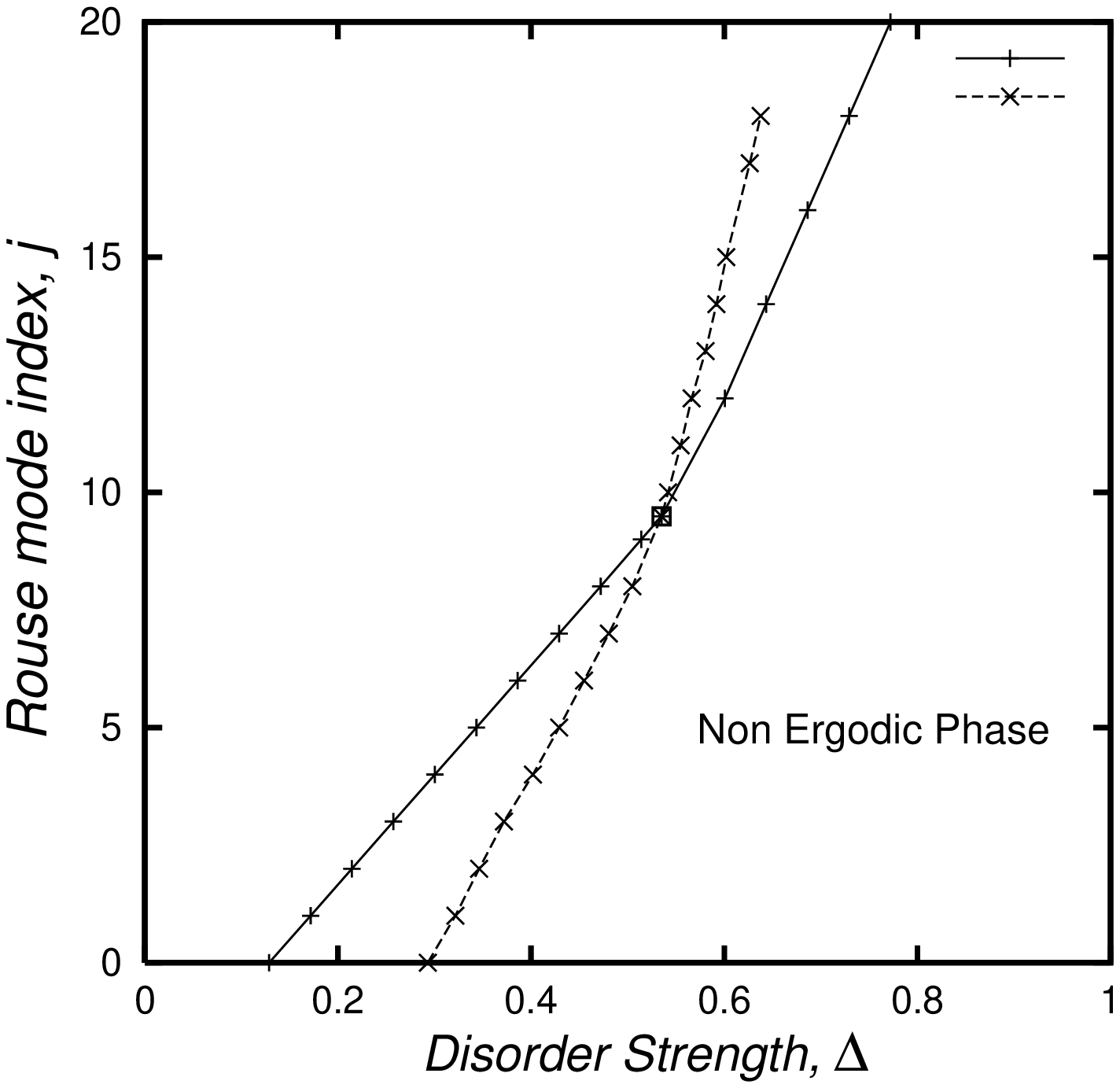}} \vspace*{10pt}
      \caption{\label{Coil_Globule} Comparison of freezing diagram for coil and globule states. Two curves correspond to two different conditions with different bare virial coefficients, namely $v = 0$ (solid line) and $v = 0.5$ (dashed line).}
  \end{minipage}
 \end{center}
\end{figure}
It is interesting to elucidate how the Rouse modes freeze in the coil state.
For this end we have driven the system to the coil state by increasing the
value of the bare virial coefficient up to $v = 0.5$ while the
eq.(\ref{Gotze_eq}) is solving. The result of the mode freezing is shown in
Fig.\ref{Coil_Globule} and is compared with the previous case (where $v = 0$).
It can be seen that the freezing of the modes in the coil state at least for
small mode indices ($0 < j < 10$) occurs at higher values of $\Delta$. This is
intuitively clear and is consistent with our finding in Sec.IV A that the
anomalous diffusion in the coil state is less affected by the disorder.

\subsection{Finite resolution and chain length study}

We have mentioned in the Sec.V A that because of very smooth bifurcation of
higher modes it is numerically not trivial to locate there critical point
$\Delta_c$. This location becomes very sensitive to the accuracy with which we
measure a first non - vanishing value of $f(p)$ while it bifurcates. We call
this resolution $h$, and the Fig.\ref{Resolution} shows that as the chain
length $N$ increases and the resolution $h$ is getting more fine (i.e. $h$
decreases) the measurable critical points of the different modes merge each other.
This proves that theoretically all modes freeze at the same critical point
$\Delta_c$. Practically, since any experiment has a finite resolution one can
presumably see that modes freeze sequentially (see Fig.\ref{3D_Diagram}).

Finally, in order to study how the critical disorder parameter scales with a
chain length we have plotted (see Fig.\ref{Finite_size}) the value of
$\Delta_{\rm c}$ for different $N$. This numerical study, which the
Fig.\ref{Finite_size} depicts, reveals that the scaling law has the form:
$\Delta_c = c N^{-\gamma}$, where the prefactor $c$ increases with the solvent
quality but $\gamma \approx 0.25$ and is universal. This shows that $\Delta_c$
for non - zero Rouse modes scales differently than for the zero - mode, which
follows the Harris freezing criterion (see the Sec. IV B). Such scaling law
obviously indicates the importance of the Rouse modes coupling.

\begin{figure}[ht]
 \begin{center}
\setlength{\unitlength}{\linewidth}  
 \begin{picture}(1.0,0.42)(0,0) 
  \put(0,0){\includegraphics[width=0.4\linewidth]{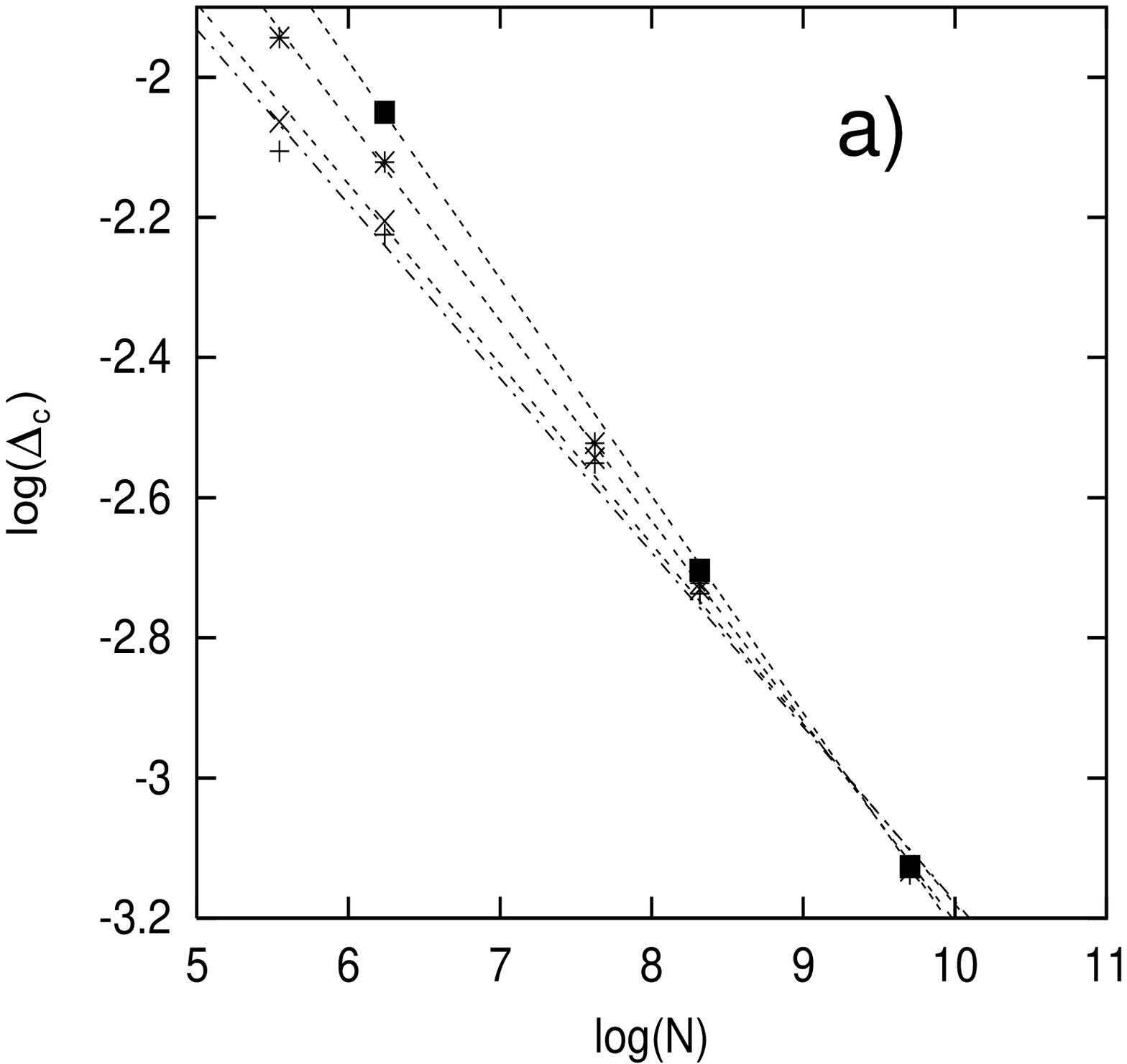}}
 \put(0.5,0){\includegraphics[width=0.4\linewidth]{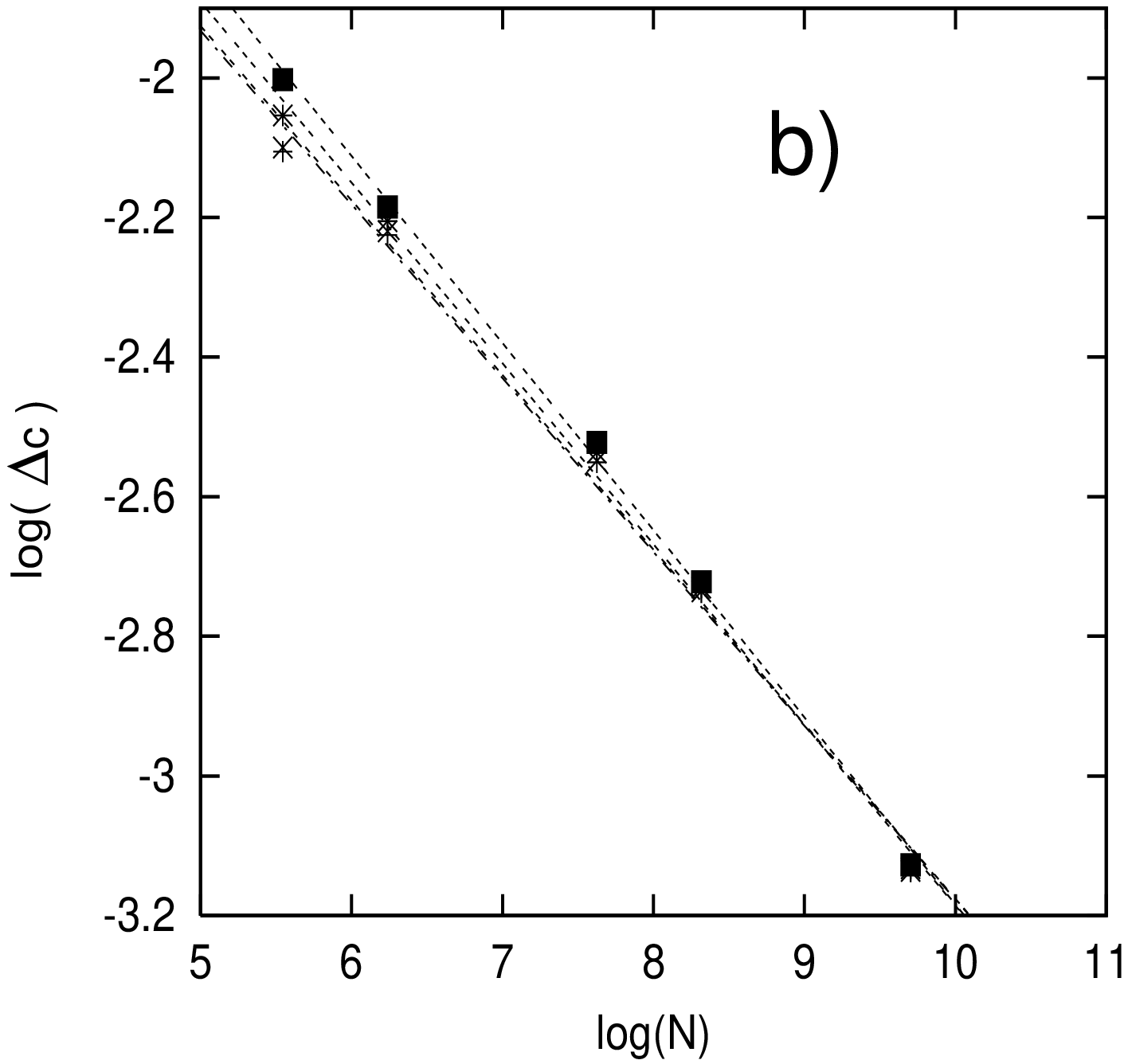}}
    \end{picture} 
\caption{\label{Resolution} $\Delta_c$ as a function of chain
  length $N$ and the resolution $h$: a) $h = 10^{-3}$, b) $h =
  6.5\times 10^{-5}$.\\ Symbols correspond to the different modes:
  $ (\blacksquare) \quad j = 1; (\divideontimes)\quad j = 2; (\times)\quad j =
  3; (+)\quad j = 4$}
  \end{center} 
\end{figure} 

\begin{figure}[ht]
  \begin{center}
  \begin{minipage}{10cm}
    \centerline{\includegraphics[width=8cm]{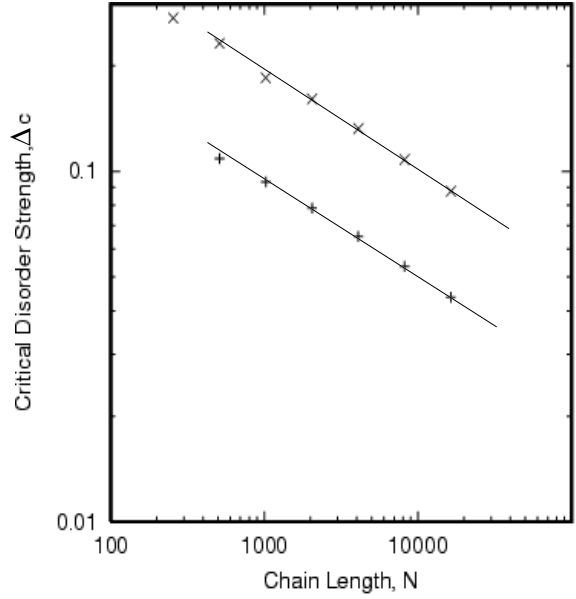}} \vspace*{10pt}
      \caption{\label{Finite_size} Finite size analysis shows how the
        critical $\Delta_{\rm c}$ for the non - zero  Rouse modes
        depends from the chain length N.Symbols correspond to the
        different solvent quality: $(+) \quad v = 0; (\times) \quad  v = 0.5$.  This reveals the scaling law
        $ \Delta_c \propto N^{-\gamma}$, where $\gamma \approx 0.25$.}
   \end{minipage}
 \end{center}
\end{figure}

\section{Conclusion}

We have studied the dynamics of the self - interacting polymer chain (with
second and third virial terms) which also experiences a quenched Gaussian
random field. Consideration is based on the MSR - generating functional method
and self - consistent Hartree approximation which allow to derive the equation
of motion for the bead-bead time - dependent correlation function
$C(s,s';t,t')$. For the time interval less then the maximal Rouse time,
$\tau_{R} \sim \tau_0 N^{1+2\nu}$, we have found an anomalous diffusion regime
and have calculated explicitly the correspondent subdiffusional exponent. At
the larger time interval, $t \gg \tau_{R}$, the ergodicity of the system is
getting broken upon the disorder strength $\Delta$ increases (i.e. the chain
becomes frozen). We have derived the corresponding equation for the non -
ergodicity function $f(p)$ and have solved it numerically. This solution shows
that with increasing $\Delta$ all modes freeze at the same critical point,
even though the higher modes bifurcate much more smooth. The center of mass
(or zero - mode) dynamical freezing is governed by the Harris criterion,
$(\Delta /b^d) N^{2 - \nu d} \ge 1$, whereas the critical point for the non -
zero modes freezing is scaled as $\Delta_c \propto N^{- \gamma}$. As it is
found numerically $\gamma \approx 0.25$ and even does not depend on the
solvent quality. This universality is presumably the result of strong mode
coupling effects. On the other hand it is of interest that the freezing line
for coil and globule states are different with the coil becomes frozen at
smaller disorder strength $\Delta$ (see Fig.\ref{Coil_Globule}).

It should be emphasized that the whole consideration in this paper is
based on the validity of FDT and time translational invariance
(TTI). It is sufficient for the  discussion of  Rouse modes freezing
close to the critical point $\Delta_c$. For the investigation of the
gyration radius freezing well below $\Delta_c$ it is necessary to
discuss ${\cal R}_g^2(t) = (1/N) \sum_{s=0}^{N-1} <~R^2(s,t)>$ in the aging
regime where FDT and TTI do not hold \cite{Bouch}. We will return to
this study  in 
following publications.

\section*{Acknowledgments}
The authors have benefited from discussions with  Matthias Fuchs and Albert Johner. V.G.R. and
T.A.V.  acknowledge financial support from the Laboratoire Europ\'een
Associ\'e (L.E.A.)

\begin{appendix}
\section{Harris criterion}

It was argued presumably for the first time by Machta \cite{Machta,Machta1}
that even though the disorder has no effect on the Flory exponent it can
influence the whole spatial distribution of the chain at $d < 4$. Before him
some authors \cite{Harris,Kim} believe that disorder is completely irrelevant
for the chain configurations provided that $v > \Delta$. In this Appendix we
paraphrase the Harris criterion argumentation given in ref.\cite{Dots} and
apply it to a self - avoiding chain problem.

Let us start from the field theory representation for a self - avoiding chain
in the presence of a quenched random field $V({\bf r})$ with the Gaussian
distribution and the second moment given by eq.(\ref{Delta}). We will use the
usual $n$ - component $\psi_{a}^4$ - field theory Hamiltonian \cite{Kholod}
which for our problem has the following form:
\begin{eqnarray}
 H = \int d^d r \Biggl\{ \frac{b^2}{2d} \sum_{a = 1}^n \left(\nabla
   \psi_{a}(r)\right)^2  + \frac{1}{2} \tau \: \sum_{a = 1}^n
 \psi_{a}^2(r) + \frac{1}{4} v \left[\sum_{a = 1}^n
   \psi_{a}^2(r)\right]^2 -\frac{1}{2} V({\bf r})\: \sum_{a = 1}^n
 \psi_{a}^2(r)\Biggr\}
 \label{field_theory}\quad,
\end{eqnarray} 
where $\psi_{a}(r)$ ($a=1,2,\dots n$) is a $n$ - component field and $\tau$ is
the chemical potential conjugated to $N$. In this representation $V({\bf r})$
acts as a random fluctuations of the effective transition temperature: $\tau
\to \tau - V({\bf r})$.

The local minima configurations are given by the saddle - point equation
\begin{eqnarray}
 - \nabla^2 \psi_{a}(r) + \left[\tau -  V({\bf r})\right]\psi_{a}(r) + v 
 \psi_{a}(r)\sum_{b = 1}^n \psi_{b}^2(r) = 0
 \label{SP_equation}\quad.
\end{eqnarray}
The Harris criterion should follow from the condition that in the average the
quenched fluctuations of $V({\bf r})$ dominate over the chain's entropy. Let
us take a region of a linear size $L$ larger then chain's size, i.e $L \gg R$.
The typical value of the frozen fluctuations in this region reads
\begin{eqnarray}
 {\overline V}_{\rm type} \sim \frac{\Delta^{1/2}}{L^{d/2}}
 \label{V_typ}\quad.
\end{eqnarray}
Similarly to ref.\cite{Kholod} let us assume that the field has ${\cal O}(n)$
internal symmetry and can be parameterized as
\begin{eqnarray}
 \psi_{a}(r) = n_{a} \phi (r)\quad,
 \label{symmetry}
\end{eqnarray}
where $\vec n$ is a unit vector in the space of the $n$ - vector model, i.e.
$\sum_{a = 1}^n n_a^2 = 1$. The vector field $\psi_a$ corresponds to the
polymer density in the following way: $\rho (r) = (1/2) \sum_{a = 1}^n
\psi_{a}^2(r) = (1/2) \phi^2 (r)$. Then the field averaged over the whole $L$
- region obeys the equation
 \begin{eqnarray}
 \tau - {\overline V}_{\rm type} + v {\overline \phi}^2 = 0
 \label{SP_equation1}\quad,
 \end{eqnarray}
 where ${\overline \phi}^2 \sim \rho$. If the fluctuations ${\overline
   V}_{\rm type}$ are strong, i.e. ${\overline V}_{\rm type} \gg \tau$, 
 the by making use of eq.(\ref{V_typ}) we get
 \begin{eqnarray}
 L \ll \frac{\Delta^{1/d}}{\tau^{2/d}} \quad.
 \label{L}
 \end{eqnarray}
 On the other hand $L \gg R \sim b \tau^{-\nu}$, where $\tau \sim 1/N$
 (i.e.$R \sim N^{\nu}$). Combining this with eq. (\ref{L}) yields
 \begin{eqnarray}
 \frac{\Delta}{b^d} \: N^{2 - \nu d} \gg 1 \quad.
 \label{Harris_once_more}
 \end{eqnarray}
 The ``specific - heat'' exponent $\alpha = 2 - d \nu  = (4 - d)/(d +
 2)$ is positive at $d < 4$ and under condition
 (\ref{Harris_once_more}) the disorder substantially affects the
 chain's statistics. This recovers the Harris criterion which we have
 discussed in Sec.II B.

 \section{Connection with $F_{\rm 12}$ - model}

 It is instructive to show how our general eq.(\ref{Gotze_eq}) is
 related with G\"otze's $F_{\rm 12}$ - model \cite{Gotz}. This
 reduction lays a solid mathematical basis for our analysis.

 First of all we will need the following orthogonality conditions which 
 can be readily obtained  by direct calculations:
 \begin{eqnarray}
 \sum_{s=0}^{N - 1} \cos (p s) = N \delta_{p0}
 \label{ortho1}
 \end{eqnarray}

 \begin{eqnarray}
 \sum_{s=0}^{N - 1} \cos (p s) \cos (q s) = \frac{1}{2} N \delta_{pq}
 \label{ortho2}
 \end{eqnarray}

 \begin{eqnarray}
 \sum_{s=0}^{N - 1} \cos (p s) \cos (q s) \cos (\kappa s) = \frac{N}{2} 
 \left\{\begin{array}{r@{\quad,\quad}l}
 \delta_{\kappa,p+q} + \delta_{\kappa,q-p} & {\rm if}\quad  q > p\\
 \delta_{\kappa,p+q} +\delta_{\kappa,p-q}  &{\rm if}\quad  q < p
 \end{array}\right.
 \label{ortho3}
 \end{eqnarray}
 Now we can expand r.h.s. of eq.(\ref{Gotze_eq}) with respect to the
 non - ergodicity function keeping in mind that close to the
 bifurcation point $f(p)$ is small. Then performing this expansion up
 to the second order and taking into consideration eqs.(\ref{ortho1}) - 
 (\ref{ortho3}) yields
 \begin{eqnarray}
 \frac{f(p)}{1 - f(p)} &=& \Delta \: \frac{d^{\frac{d}{2} + 1}(d + 2)}{2^{d + 3}
   \pi^{\frac{d}{2}}} \: \frac{N^2 C_{\rm st}^2(p)}{\left[\sum_{q =
     2\pi/N}^{2\pi} C_{\rm st}(q)\right]^{\frac{d}{2} + 2}} \Biggl\{
 f(p) + \frac{d + 4}{8 C_{\rm st}(p)\left[\sum_{q =
     2\pi/N}^{2\pi} C_{\rm st}(q)\right]}\nonumber\\
 &\times& \Bigl[\sum_{q = 2\pi/N}^{2\pi} C_{\rm st}(q) C_{\rm
   st}(p+q)f(q)f(p+q)  + \mathop{{\sum}'}_{q = 2\pi/N}^{2\pi} C_{\rm
   st}(q)C_{\rm st}(|p-q|) f(q) f(|p - q|)\Bigr]\Biggr\} \quad,
 \label{F_12}
 \end{eqnarray}
 where $\mathop{{\sum}'}_{q = 2\pi/N}^{2\pi}$ means that term $p = q$ is
 dropped out. It is of interest that because of eq.(\ref{ortho1}) the zero -
 order term in this expansion is vanished, so that eq.(\ref{F_12}) always has
 the trivial solution $f(p) = 0$. The eq.(\ref{ortho1}) has the functional
 structure of $F_{\rm 12}$ model which has been studied by G\"otze
 \cite{Gotz}.

\end{appendix}

\end{document}